# Investigating intrinsic silicon oxide and a lightly doped p-type layer for evolution of tail states and estimating its impact as a passivation layer in a HIT type device

S M Iftiquar[1,2]

[1] Indian Association for the Cultivation of Sciences, Raja S. C. Mullick Road, Kolkata 700032, India

[2] College of Information and Communication Engineering, Sungkyunkwan University, South Korea

Abstract

   Hydrogenated amorphous silicon is well known for its various alloys and wide ranging opto-electronic properties. Hydrogenated silicon sub-oxide (aSiO:H) is one of them. The effect of boron doping on optoelectronic properties of the aSiO:H have been investigated with intrinsic and delta-doped materials. The Urbach energy of the tail and midgap defect states of the intrinsic and lightly doped(0.01% gas phase doping) materials were found to be 103, 151 meV and $8.19 \times 10^{16}$, $5.47 \times 10^{17}$ cm$^{-3}$ respectively. As a result of the 0.01% doping a reduction in optical gap was observed from 1.99 to 1.967 eV along with a shift in Fermi level by 0.258eV, indicating an efficient boron doping of the intrinsic silicon oxide material. The lightly doped p-type layer was used to simulate device characteristics in amorphous silicon (aSi:H) and HIT-type solar cell. In the aSi:H device the power conversion efficiency (PCE) was moderate while in a HIT-type device the PCE was 26.4%. It indicated that a good quality silicon oxide layer can be an attractive passivation layer for a HIT type device. It further indicated that defect density of the passivation layer can have a significant impact on performance of a HIT type device. The primary difference of the two passivation layers (intrinsic and delta-doped) are the difference in mid-gap and tail states. Passivation layer with a higher mid-gap state leads to poorer device performance. Therefore, it highlights the importance of using a less defective passivation layer to improve device performance.

**Keywords:** Hydrogenated amorphous silicon oxide; Boron doping; Constant Photocurrent Method; Urbach Energy; Defect density; Heterojunction solar cell

**Corresponding Author**: smiftiquar@gmail.com

## 1.     Introduction

   In silicon devices the heterojunction structure is used to enhance the desired unidiretionality of the flow of photo-generated charge. The directional charge flow is achieved by two principal methods, by using doped layers and heterogeneous structure of energy band. For silicon devices, various silicon alloys have been widely investigated and reported by various research groups. One of such wide band gap material is hydrogenated silicon sub-oxide (aSiO:H). With the addition of small amount of oxygen to amorphous silicon (aSi:H) network, its optical gap ($E_g$) can be enhanced keeping its electronic properties nearly unchanged. However, it was noticed that with an increased addition of oxygen to the aSi:H, its optical gap steadily increases and its electrical conductivity decreases [1], although such changes depend upon various factors [1, 2]. Considering this as a natural and unavoidable consequence of adding oxygen to the aSi:H, its better to select or optimize the material with highest possible optical gap and with lowest possible reduction in electrical conductivity.

The aSiO:H is one of the many of the silicon alloy investigated for its application in solar cell [3-6], because its optical gap is wider [1, 2] than its parent aSi:H material and with a relatively low electronic defect density [1, 2]. Characteristics of intrinsic and doped silicon oxide has been reported [1, 2, 7, 8].

Doping in amorphous silicon is generally carried out in the gas phase, during the material or film deposition. This method of doping is quicker and easier but it carries a risk of deterioration of the material if the atomic bonding within the material gets significantly altered due to the doping process. Increased defect leads to a reduced effect of doping. It seems this is the case because in most of the reported results the increased boron doping does not reduce the activation energy as expected.

Dopability of the amorphous silicon alloy depends on its midgap defect density; at a higher defect the Fermi level remains pinned and does not move significantly with increased level of dopant atoms. In the year 1969, Chittick et al. [9] reported preparation of photosensitive amorphous silicon from silane source gas. Soon afterwards it was realized that defects in amorphous silicon is significantly reduced by the bonded hydrogen atoms, thereby making the hydrogenated amorphous silicon as photo sensitive and dopable. These defects can be called as midgap defects as it exists within the valence and conduction band energy levels. At this high level of midgap defect, the electronic transitions are difficult to detect optically. However, the subgap photon absorption can lead to a certain level of electrical conductivity [10]. This electrical conduction can be detected and used as a signature of subgap absorption, hence midgap defect density can be estimated from such a photon assisted electrical transport. One of the methods that can be used to estimate the midgap defect density is constant photocurrent method (CPM) [10, 11]. In this method, a variable wavelength radiation is incident on the sample. The photocurrent to the external circuit is kept constant by varying intensity of incident light. In this sub band gap illumination, light excites electron from the valence band and midgap defects get populated. The trapped electron from the defective cites can get excited further by the incident radiation, leading to current in the external circuit. As such a photo-current depends upon defect density, so the defect density can be estimated by the CPM measurement. Furthermore, the Urbach tail state that exists between the extended valence and conduction band states indicates the level of disorder in atomic bonding within the amorphous silicon network. With the help of the CPM these states can also be explored. As the CPM depends upon photosensitivity of the material, so it is difficult to characterize the midgap states of a highly doped material, as highly doped materials are not significantly photo sensitive.

Therefore, a lightly doped p-type silicon oxide (p-aSiO:H) material was used to investigate the evolution of the midgap defect state with reference to an intrinsic silicon oxide material (i-aSiO:H). This material was used in silicon heterojunction (HIT-type) solar cell simulation, by using AFORS-HET simulation program. The simulation result shows that a lightly doped p-aSiO:H may be useful for high efficiency HIT type solar cell.

## 2. Theory

### 2.1. Photoconductivity

Enhancement in electrical conductivity due to light incidence on semiconductor materials is called photo–conductivity, $\sigma_{ph}$. Excess electron hole pairs are generated by the incident light that is absorbed by the photosensitive layer. In a steady state condition, if n and p be the average density of excess electron and hole pair generated per unit volume within the thin film due to the incident light, then the primary photo–conductivity ($\sigma_{ph}$ ) can be written as,

$$\sigma_{ph} = e( n\mu_n + p\mu_p )$$ ( 1 )

where, $\mu_n$ and $\mu_p$ are mobility, n and p are number density of electron and hole respectively.. These electrons and holes, while present in a sea of charge carriers and defects, will have an average lifetime ($\tau$) to exist as a free carrier, then in a steady state

$$n = G\tau_e \qquad\qquad p = G\tau_h$$ ( 2 )

where, $\tau_e$ , $\tau_h$ are mean lifetime for electrons and holes, G is generation rate of the electron hole pair. The G is dependent on total incident photon flux, as well as absorption capability of the material or optical absorption coefficient $\alpha$. Other than absorption, a part of the incident light gets reflected and transmitted. Generally for electron and hole the ratio of mobility is $\mu_n/\mu_p > 10$ , then a significant part of electrical conduction takes place due to the electrons. In that case, if the effect of holes is negligible, the phoconductivity may be approximated as

$$\sigma_{ph} \cong en\mu_n$$ ( 3 )

And it can be called secondary photoconductivity. For monochromatic incidence light G may be expressed as

$$G = \eta N_\lambda \ ( \ 1-R \ )[1 - \exp(-\alpha_\lambda d \ )] \qquad (4 \ )$$

where, $\eta$ is the quantum efficiency of generating free carriers for each absorbed photon. $N_\lambda$ is the incident photon flux at wavelength $\lambda$, R coefficient of reflection at the sample surface to which light is incident, $\alpha_\lambda$ absorption coefficient at a particular incident wavelength ($\lambda$) and d sample thickness. The resulting monochromatic photoconductivity can be expressed as

$$\sigma_{p\lambda} = eN_\lambda \ ( \ 1-R \ )\eta\mu_e\tau_e\{1 - \exp(-\alpha_\lambda d \ )] \qquad ( \ 5 \ )$$

It is to be noted that the expression (4), (5) has thickness dependence for a thin sample, or depth dependence if infinitesimal layers are considered for the film.

In an intrinsic photosensitive material, electron-hole pairs are generated by incident light. In an amorphous silicon alloy, the defect mediated recombination rate is higher than the band to band transition; therefore reducing the midgap defect is one of the primary research interests in developing high quality material.

## 2.2. Constant Photoconductivity

In the CPM, the subgap absorption coefficients are measured by illuminating the sample with sub-gap energy photon under the condition of constant photo-current [10, 11]. In this method it is assumed that the quantum efficiency, mobility and carrier lifetime are constants for the range of excitation wavelength used, therefore, in a low absorption region where sub-band gap photons are used to generate photo current, the absorption coefficient will be inversely proportional to the required incident photon flux for the constant current [10]

$$\alpha_\lambda = constant/N_\lambda \qquad (6)$$

Although the above equation should give absolute values of the absorption coefficient, yet due to the approximation between the expressions (5), (6), due to the assumed constancy of $\eta\mu_e\tau_e$ and systematic error in estimating the $N_\lambda$, the CPM absorption spectra is normalized with the help of band-to band optically measured absorption coefficient. In our sample that optical absorption coefficient was obtained by ultraviolet visible (UV-Vis) spectrometry.

The subgap absorption spectra have two distinct zones, the Urbach exponential tail region and a relatively flat midgap defect region. The midgap defect density can be estimated by using the formula [12]

$$N_d = 3.7 \times 10^{16} \times \alpha_{1.2} \text{ cm}^{-3} \tag{7}$$

This expression (7) is primarily formulated for hydrogenated amorphous silicon [12] yet it may be used in intrinsic hydrogenated amorphous silicon oxide for a comparative study [2]. But in case of extrinsic or highly doped semiconductor, where the charge is carried by majority carriers (electrons in n-type semiconductor or holes in p-type semiconductor), the effect midgap defect density on the carrier transport may not be significant, although in a way the defect density determines efficiency of doping. Furthermore, for a lightly doped semiconductor, when its phoconductivity is measurable, this formula may be used to gain an understanding about the possible defect density for the doped material as well [11]. When the light is incident in the doped materials, excess electron-hole pairs will be generated. But the effect of light on the enhancement of the conductivity will depend upon level of doping. The photo generated majority carrier density (hole density for p-type material and vice versa) is expected to be low compared to its existing density and the minority carriers (e.g. electrons for p-type material) may face recombination (with excess holes present in the p-type material). As a result the observed enhancement in photo current or photo conductivity will be low. Therefore, for a highly doped sample the CPM measurement is not generally possible.

Furthermore, as the recombination of the minority carriers (here electrons) is expected to be mostly taking place with the existing majority carriers (excess holes). So the mid gap states revealed by the CPM measurement of a p-type material is expected to be mostly of holes.

## 2.2. Device Simulation

In case of applying the material to a reference HIT solar cell, we used n-type silicon wafer as the absorber layer with intrinsic 7 nm thick aSi:H passivation layer to both the sides. Two types of devices were simulated, one is amorphous silicon solar cell with p-type silicon oxide layer at the front surface and the other is HIT-type device. In aSi:H solar cell the low doped p-layer was used in place of the standard 1% doped one. For the HIT-type device the cSi/p-aSi:H interface passivation was changed from intrinsic aSi:H to 0.01% doped aSiO:H layer (δp-aSiO:H) layer and an intrinsic aSiO:H layer.

# 3.    Experimental

## 3.1.    Film Deposition

The sample was deposited by 13.56 MHz RF plasma enhanced chemical vapor deposition system on non-textured Corning glass for optical and electrical characterization. For Fourier transform infrared spectroscopy the samples were deposited on crystalline silicon wafers. The silane ($SiH_4$), carbon dioxide ($CO_2$), hydrogen ($H_2$) source gases were used to prepare intrinsic aSiO:H film while an additional $B_2H_6$ gas was used for its doping. The deposition condition used in preparing the films is given in Table I

The $1.5 \times 1.5$ cm$^2$ films were deposited with a $1.5 \times 1.5$ cm$^2$ open area metal mask. Thicknesses of these films were measured by using stylus type instrument with a sensitivity of 15 nm. Thicknesses of the deposited films were kept around 600 nm so that error in measuring thickness and optoelectronic properties are low. Coplanar aluminum electrodes are deposited on the rectangular sample surface by thermal evaporation. Separation of the electrodes were 0.1 cm. Electrical connections are taken from these electrodes to measure electrical conductivity.

## 3.2.    Electrical conductivity

Dark and photo conductivities were measured in a vacuum chamber (cryostat), at a pressure of $10^{-6}$ Torr obtained by an oil diffusion pump backed by a rotary pump. Prior to measuring the temperature dependence of conductivity, each sample is annealed at ~150 $^o$C temperature for an hour. Temperature was monitored by a temperature controller with the help of copper constantan thermocouple. Dark and photo currents were noted at 100 V applied DC bias. Amorphous silicon and aluminum junction will form a metal semiconductor Schottky barrier. In order to avoid the diode characteristics during electrical measurements the linearity of the junctions is tested with varying applied bias and it is found that 100 Volts to lie within the linear part corresponding to an electric field of $10^3$ V/cm. Current is measured by using Kiethley electrometer. In order to measure photo conductivity of the material, a 100 mW/cm$^2$ intensity white light was incident on the sample through a quartz window. Before measuring the photoconductivity, the reference 100 mW/cm$^2$ light intensity was set by measuring the light intensity at the sample site with the help of a standard silicon photodiode, placed at the sample position inside the cryostat. Its dark conductivity can be expressed as

$$\sigma_d = \sigma_o \exp[-\Delta E/kT ) \qquad (8)$$

Where the $\sigma_o$ is a material specific constant, $\Delta E$ is activation energy which is the energy difference between the Fermi level and mobility edge, k is Boltzmann constant and T is sample temperature.

## 3.3.    FTIR

In order to measure hydrogen and oxygen bonding to the Si atoms of the deposited films, Fourier transform infrared (FTIR) spectrometer was used (Perkin Elmer FTIR spectrophotometer). The effect of c-Si substrate on the FTIR spectra is minimized by correcting with a reference FTIR spectra from a similar c-Si wafer. The O atoms bonded in the aSi network will exhibit a characteristic FTIR absorption spectra at around 980 cm$^{-1}$ along with a modification of Si-H vibrational spectra around 2000 cm$^{-1}$ [2, 13, 14].

## 3.4.    Optical Absorption

The optical gap (Eg) has been estimated from optical absorption ($\alpha$) spectra obtained from a UV–visible spectrophotometer, in the 2.0–2.6 eV energy region, using Tauc's formula:

$$(\alpha h\nu)^{1/2} = B (h\nu - Eg ) \qquad (9)$$

Here h is Plank's constant, $\nu$ is optical frequency, B is another constant.

**Table-I.** Deposition condition and properties of the films. The gas flow rates are $SiH_4$ ,$H_2$, $CO_2$, as 10, 82.5, 7.5 sccm respectively, while RF power density 30 mW/cm$^2$, pressure 0.8 Torr, Substrate temperature 200 $^o$C. Here optical gap was estimated by using Tauc's relation.

| Sample | $B_2H_6$ (sccm) | Optical Gap, $E_g$ (eV) |
|---|---|---|
| i-aSiO:H | 0 | 1.99 |
| p-aSiO:H | 0.0001 | 1.97 |
| p-aSiO:H | 0.0005 | 1.97 |
| p-aSiO:H | 0.002 | 1.93 |
| p-aSiO:H | 0.008 | 1.81 |
| p-aSiO:H | 0.01 | 1.79 |

## 3.5.    CPM

CPM measurements have been carried out under $10^3$ V cm$^{-1}$ electrical drift field that was kept fixed during the measurements. The measurements were performed in a vacuum ($\sim 10^{-3}$ Torr) chamber. For the monochromatic light source a tungsten halogen lamp was used along with concave mirror and a focusing convex lens to guide high light output towards the sample. A monocromator was used to select radiation wavelength for the sample illumination. Intensity of the monochromatic light was calibrated at each wavelength by a silicon photodiode. The constant electrical current (and hence electrical conductivity ) was monitored by giving enough time to stabilize the current reading and was controlled by regulating electrical power to the halogen lamp. In the CPM method, this current was kept constant.

# 4.        Results and Discussions

The measured experimental results and discussions are given below. Here our investigation is primarily on the tail states and mid gap states of the lightly boron doped p-type aSiO:H films. The CPM measurements were not possible for other doped samples. This was because the limitation of sensitivity of the CPM measurement system. Furthermore we investigated the effect of lightly doped sample as a passivation layer at the p-type emitter side. So our investigation is primarily based on the intrinsic aSiO:H and 0.01% boron doped films. The trend in optoelectronic properties of the other doped samples are similar to other films reported by other research groups. It can be described in brief as the following. The optical band gap, Eg, decreased steadily from 1.99 eV (intrinsic) to 1.79 eV for highest doping while dark conductivity increased to $2\times10^{-5}$ S.cm$^{-1}$ for the highest doping.

## 4.1.    Optoelectronic Characteristics

Figure 1(a) shows the optical absorption spectra for the intrinsic i-aSiO:H  and 0.01% doped p-aSiO:H films. The optical absorption coefficient increases with the 0.01% doping, implying that the optical gap of the doped film will be lower than the intrinsic film. Tauc gap ($E_g$) of the 0.01% doped and undoped films was estimated to be 1.967 and 1.99 eV respectively, as shown in Table II. Reduction in the optical gap with boron doping of amorphous silicon alloy is well known [15]. The reason could be an additional disorder of the amorphous silicon network introduced by the boron doping. With an increased doping, the optical gap of silicon oxide material decreases further .

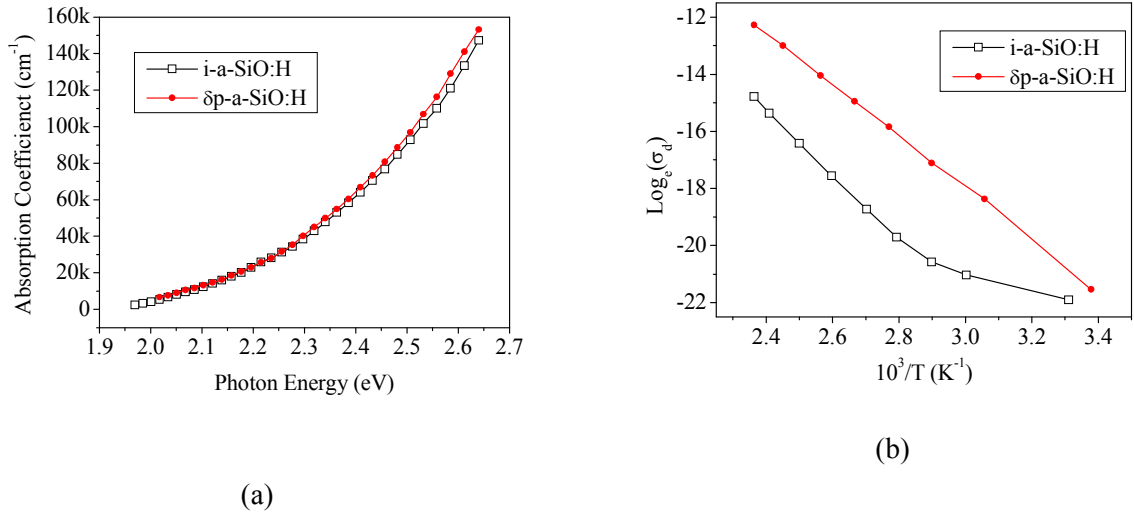

(a)

(b)

**Fig. 1.** (a) Optical absorption spectra of the intrinsic and doped films obtained by UV-Vis spectrometric measurement. (b) $Log_e(\sigma_d)$ vs $10^3/T$ Arrhenius plot for the same samples.

Primary motivation of p-type doping of the silicon oxide film is to shift its Fermi level close to the valence band (VB) density of states (DOS). One indication about the relative distance between the VB and the Fermi level is to estimate the activation energy, $\Delta E$. As the dark conductivity depends upon $\Delta E$ and temperature, T, by varying the temperature and measuring the dark conductivity the $\Delta E$ was estimated from the slope of the Arrhenius plot. Higher the slope indicates higher $\Delta E$. The estimated activation energy ($\Delta E$), as measured for the changes from 0.948 to 0.784 eV for the intrinsic and p-type samples having optical gap of 1.99 and 1.967 eV respectively. Here the midgap energy ($E_{mid}$) for the intrinsic and doped materials are 0.995 and 0.9335 eV respectively. As the intrinsic materials are slightly n-type because electron has higher mobility than that of the hole, therefore, its activation energy ($\Delta E_i$) is effective energy difference between the Fermi level ($E_F$) and conduction band mobility edge ($E_{CB}$):

$$\Delta E_i = E_{CB} - E_F \tag{10}$$

The numerical values indicate that the Fermi level for the intrinsic material is close to the midgap energy level. On the other hand the when this material is doped with boron, an intentional addition of hole makes this material p-type as hole becomes majority charge carrier. The Fermi level crossed the $E_{mid}$ and moved towards the VB. Therefore, the activation energy correspond to the difference in energy from Fermi level to the valence band mobility edge ($E_{VB}$).

$$\Delta E_p = E_F - E_{CB} \tag{11}$$

In the intrinsic material the energy difference between the $E_F$ and $E_{CB}$ was estimated to be 1.042 eV (which is $E_g - \Delta E_i$ ), whereas for the p-type material it is directly measured as 0.784 eV. This implies a shift in Fermi level by 258 meV due to 0.01% doping. However, the shift in the Fermi level with increasing the doping is not expected to be linear, meaning, a proportional shift is not expected to happen at a higher doping level, otherwise with a 1% or 2% doping a shift in the Fermi level by 18.2 or 36.4 eV could have been achieved or the material could have been degenerate, which obviously was not reported .

The reduction in the activation energy is (0.948-0.784 or) 0.164 eV which is smaller than expected for a solar cell device application, but shift in the Fermi level was 258 meV. This low doping ca maintain a level of photo conductivity necessary for the CPM measurement. At the same time it is also expected to alter the valence band tail state that can be detected by the CPM. At this stage the intrinsic and p-type films show photo conductivities ($\sigma_{ph}$ ) of $1.4 \times 10^{-6}$ S.cm$^{-1}$ and $2.2 \times 10^{-7}$ S.cm$^{-1}$ whereas its dark conductivities were ($\sigma_d$ ) of $8.0 \times 10^{-12}$ S.cm$^{-1}$ and $7.4 \times 10^{-10}$ S.cm$^{-1}$ respectively. Here the photo conductivity is the change in electrical conductivity due to 100 mW/cm2 intensity white light incident of the film. The significant reduction in the $\sigma_{ph}$ and increase in the $\sigma_d$ indicates that photo conductivity degrades along with increase in dark conductivity due to the boron doping.

Here activation energy is estimated from slope of the plot, but the two plots have a relative vertical shift. So the conductivities are different; doped film has higher conductivity than the undoped one. Furthermore, the electrical conductivity (and mobility) measured in a planar electrode configuration, with 1 mm inter-electrode spacing, may be different from that measured across the few nanometer thick film, perpendicular to the film surface surface, like that during operation of a solar cell. Because of statistical collision of carriers during 1 mm transit the effective mobility is expected to be lower than that across few nanometer thick film where the statistical fluctuations will be and the carrier transit will most likely be a ballistic type. Therefore, we assume a high values of charge mobilities during device simulation, as given in Table II.

## 3.1. FTIR

Figure 2 shows the FTIR spectra of the films from 550 to 2500 cm$^{-1}$ wave number. The Si-H and Si-O-Si related absorption peaks are shown in the figure. Here it can be seen that Si-H related infrared (IR) absorption decreases along with increase in oxygen related absorption, implying that hydrogen content of the film decreases whereas oxygen content increases. Even though the oxygen content within the film increases yet its optical gap remains marginally lower, because, although increased O atom within the amorphous silicon network helps to increase its optical gap yet increased boron content and reduced hydrogen content are the reasons for the observed reduction in the optical gap. At a higher doping level a further reduction in the optical band gap is expected .

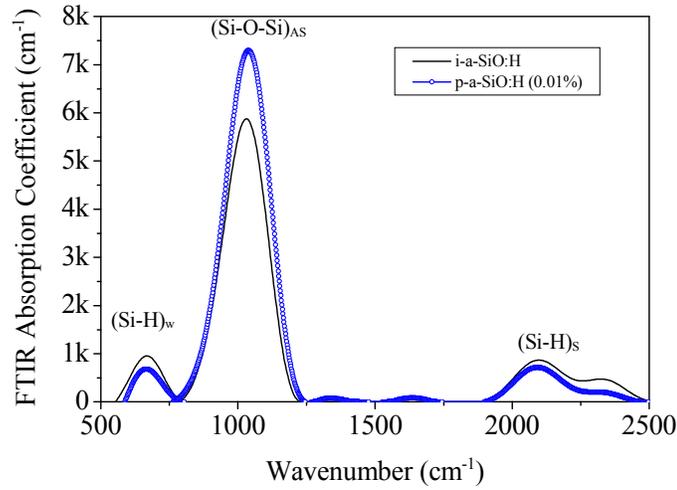

Fig. 2. FTIR spectra of the intrinsic and p-type doped samples.

## 3.2. Tail states

The band tail states were obtained with CPM measurements at room temperature, and are shown in Fig. 3. The absorption spectra can also be called as electrical conduction related optical absorption, as the optical absorptions are estimated from measuring the photon flux density at the constant current. The CPM absorption spectra is usually calibrated with optical absorption spectra, so here the UV-Vis absorption spectra was used to normalize the absorption coefficient. Based on this, the Urbach tail and defective regions are utilized to characterize the films. The inverse slope of the exponential absorption tail ($E_u$) is used as a measure of the Urbach energy ($E_u$):

$$\alpha = \alpha_0 \exp\left(\frac{h\nu}{E_u}\right) \tag{12}$$

Here $\alpha_0$ is the constant prefactor. The $E_u$ of the intrinsic film is estimated to be 103 meV whereas after 0.01% doping it became 151 meV. In Fig. 3, the flattening of the band tail state implies increase in the $E_u$. As a result the midgap density of states also increase, as shown in Table II. The reduction in the photoconductivity can also be due to increased $E_u$ and mid gap states. With an increased boron doping, the hole density in the material increases. The p-type materials are generally used for charge extraction in solar cell, and the increased hole density will helpful to achieve higher open circuit voltage ($V_{oc}$) and improved device performance.

As noticed in the Fig.3 and Table II, the Urbach tail becomes broader with Boron doping, implying an increased disorder in amorphous silicon oxide network. The rise in defect density also implies that expected movement of the Fermi level will become lower because of the tail state as well as midgap defects. If this trend remain with a further increase in doping, the movement of the Fermi level towards the valence band will be lower than that expected from the observed effect at low level doping. The trend of reduction in its optical gap with boron doping is consistent with that reported earlier .

In a solar cell, a thin p-type window layer is used through which light can transmit to the intrinsic layer. The optical absorption in the intrinsic layer generates the electron-hole pairs. The holes move towards the p-type layer while the electrons move towards the n-type layer, thereby the absorbed light is converted to electrical energy. However, it is to be noted that optical absorption in the p-type window layer in a p-i-n type solar cell is undesirable. Therefore, a thin and optically transparent p-type window layer is generally used.

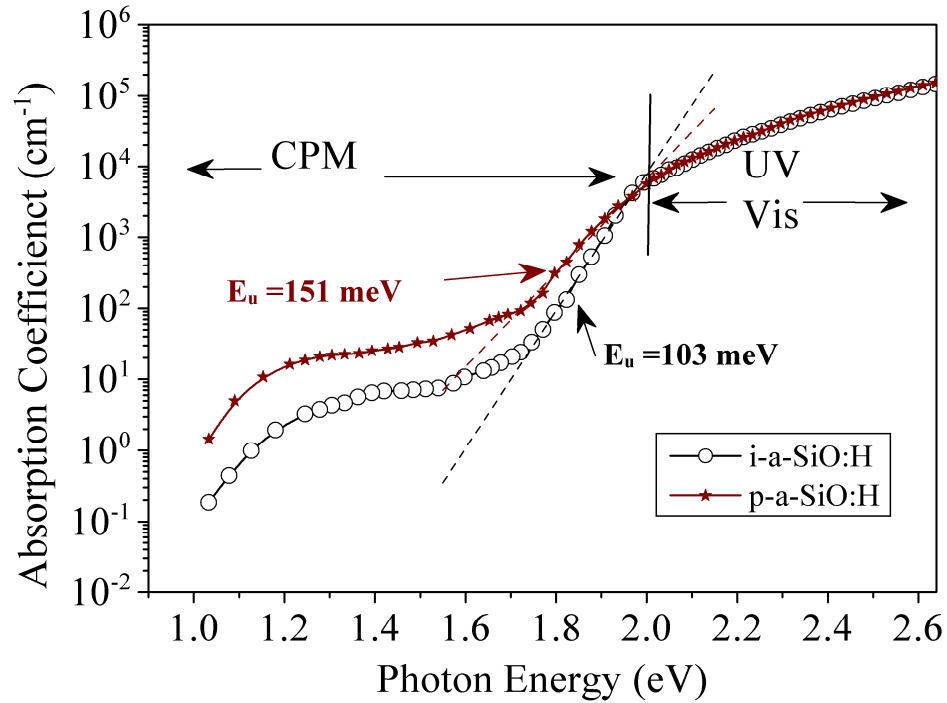

**Fig. 3.** CPM absorption spectra of the intrinsic and p-type films.

**Table-II.** Opto-electronic properties of two of the films. Here $E_g$ optical gap, $\sigma_{ph}$ photo conductivity, $\sigma_d$ dark conductivity, $\Delta E$ activation energy, $N_d$ mid gap states or defect density. For (1% boron doped) p-aSiO:H, the $E_u$, $N_d$ could not be measured.

| Sample | $E_g$ (eV) | $\sigma_{ph}$ (S.cm$^{-1}$) | $\sigma_d$ (S.cm$^{-1}$) | $\Delta E$ (eV) | $E_u$ (meV) | $N_d$ (cm$^{-3}$) |
|---|---|---|---|---|---|---|
| i-aSiO:H | 1.99 | $1.4 \times 10^{-6}$ | $8.0 \times 10^{-12}$ | 0.948 | 103 | $8.19 \times 10^{16}$ |
| $\delta$p-aSiO:H | 1.967 | $2.2 \times 10^{-7}$ | $7.4 \times 10^{-10}$ | 0.784 | 151 | $5.47 \times 10^{17}$ |
| p-aSiO:H | 1.79 | - | $2.0 \times 10^{-6}$ | 0.540 | - | - |

In order to estimate the effect of doping on the loss or absorption of incident light at the p-type layer, its optical absorptions were estimated. The standard 100mW/cm$^2$ intensity AM1.5G was used to estimate the absorbed light at the p-type layers.

### 3.3. Parasitic Optical Absorption

Parasitic optical absorption is the undesired loss of light due to absorption at the p-type window layer. With an increased doping although the Fermi level comes closer to the valence band but the optical transmission reduces. Both these trends have opposite effect on device performance. Therefore, a high level of doping may not result in a high device performance. Figure 4 shows the optical absorption spectra of the layer. The optical absorption was estimated by using Beer Lambert's exponential relation for intensity of optical transmission ($I_T$).

$$I_T = I_0(1-R)\exp(-\alpha d) \tag{13}$$

Here $I_0$ is intensity of incident radiation, $\alpha$ absorption coefficient, d is film thickness. The upper trace in the Fig. 4 is for the 7 nm thick p-aSiO:H, which shows a small increase in absorption with reference to the intrinsic layer. It can also be noted that the rise in optical absorption are mainly in the extended state, and not significantly different near the tail state or midgap state; because absorption coefficient in these regions is very low. However, the tail state and the midgap state controls the characteristic electronic properties of the material.

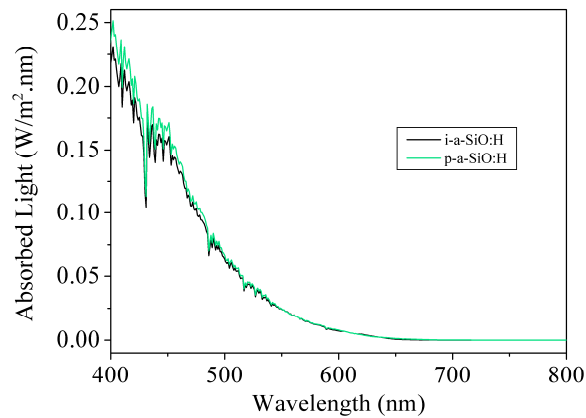

**Fig. 4.** Optical absorption of incident AM1.5G radiation by the i-aSiO:H and p-aSiO:H layer in the solar cell structure (without textured surface).

### 3.4. Simulated Device

This p-aSiO:H layer was used to simulate characteristics of a solar cell in a p-i-n type amorphous silicon solar cell with device structure, with ITO(80 nm)/ p-aSiO:H (7nm)/ i-Si:H(300nm)/ n-aSi:H(25nm) [16]. A HIT-type solar cell was also simulated. In both the cases the effect of using the 0.01% doped p-aSiO:H layer was investigated with the reference cell. For aSi:H cell the reference was created with 1% doped p-layer, and for the HIT-type cell the reference was created with aSi:H passivation layer.

In the aSi:H cell, the 0.01% doped p-aSiO:H layer was used as sole p-type layer whie in the HIT-type cell the cSi(n)/p-aSiO:H interface passivation was used with the 0.01% doped layer.

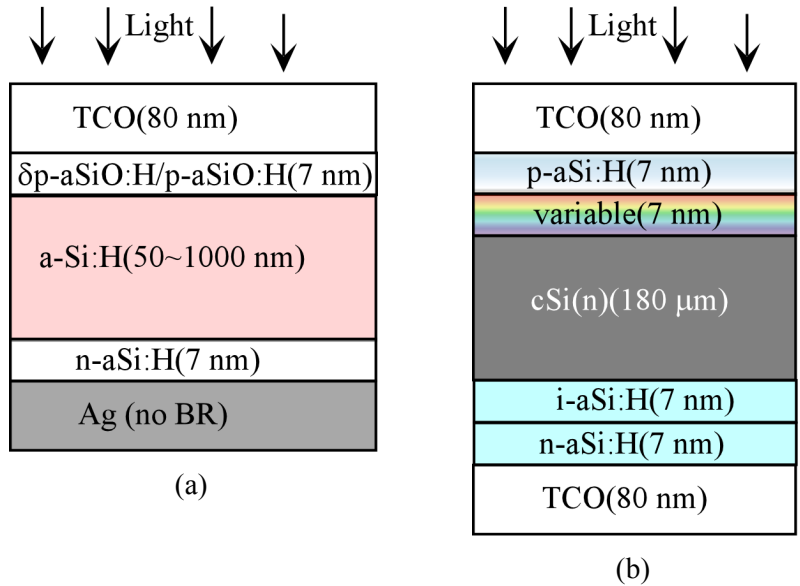

(a)

(b)

**Fig. 5.** Schematic diagram of devices used in the simulation for (a) amorphous silicon solar cell, (b) HIT solar cell.

The few of the optoelectronic simulation parameters are given in Table II & the device structures used for the simulation are shown in Fig. 5.

## 3.5.    Amorphous silicon solar cell

AFORS HET [17] simulation program was used for simulating current density vs voltage (J-V) characteristic curves of the heterojunction solar cells. For amorphous silicon solar cell (ASSC) the two different p-layers were used, one with 1% doping (Cell-A) and the other with 0.01% doping (Cell-B). With the 1% doped p-layer the built-in electric field is expected to be higher than that with 0.01% doped p-layer, as a result the J-V curves were different. The Cell-A show relatively lower open circuit voltage ($V_{oc}$), short circuit current density ($J_{sc}$), fill factor (FF), current density at maximum power (PmaxJ), voltage at maximum power point (PmaxV) and power conversion efficiency (PCE). The device parameters are shown in Fig. 6. It clearly shows that the FF and PCE of the Cell-A is much better than that of Cell-B. It also shows that the maximum PCE was visible at 300 nm thick active layer for both the cells.

Here simulated device characteristics is better than that generally reported for experimentally measured devices, may be primarily because the defect densities used here are lower than that happens in actually fabricated devices. In a device fabrication multiple layers are prepared sequentially. This may introduce the additional defects that are minimum in purely single layer that we used here.

Here the simulated devices are not optimized, so the reference device parameters will not be among the reported high values. But the same device structure was used for both the reference as well as new device with 0.01% doped silicon oxide passivation layer and the device characteristics were compared.

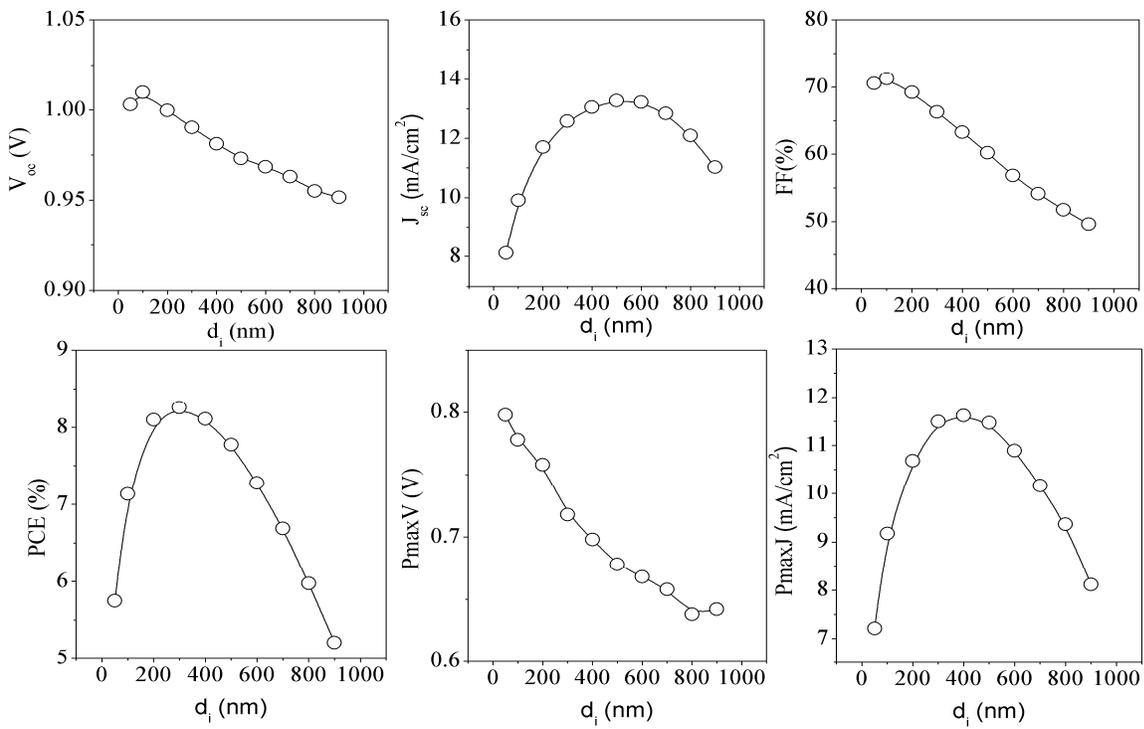

(a)

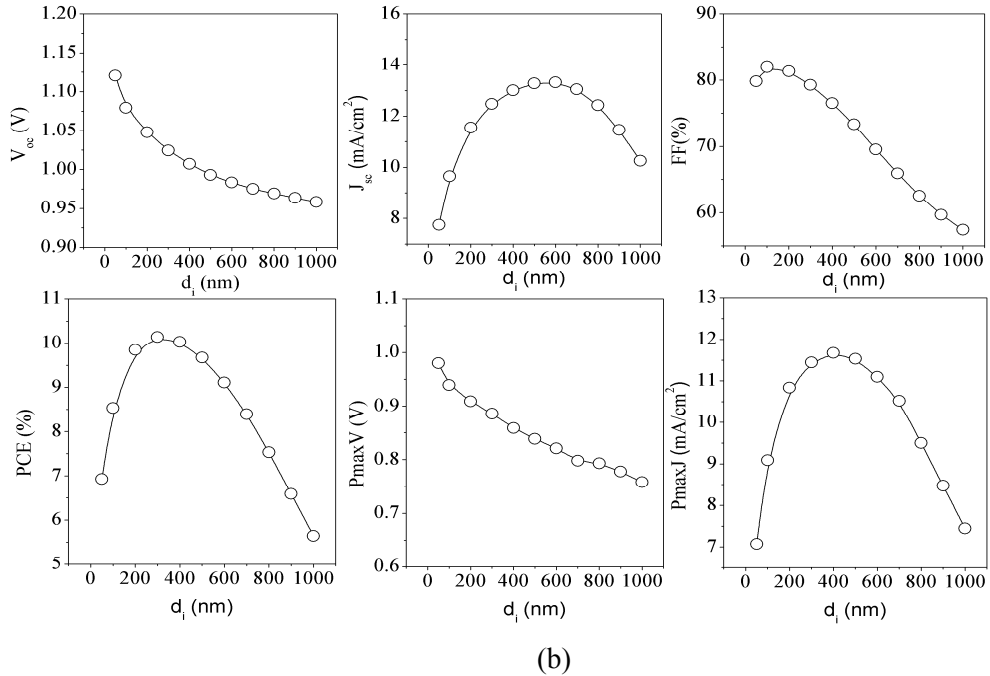

(b)

**Fig. 6.** Amorphous silicon solar cell parameters obtained with (a) 0.01%, (b) 1.0% doped p-aSiO:H. In the simulation study the active layer thickness was varied.

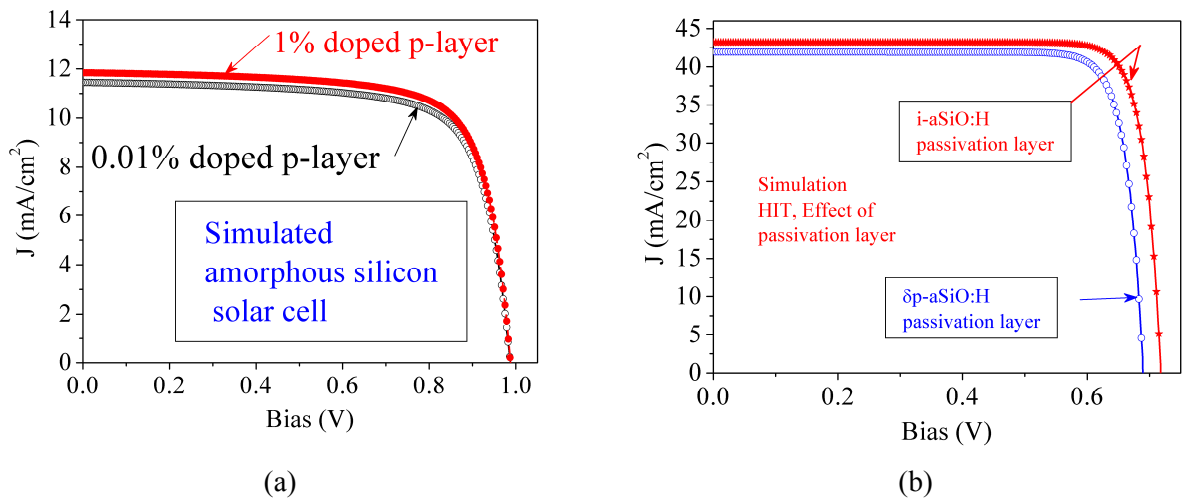

(a)                                                                    (b)

**Fig. 7.** J-V characteristic curves for (a) 300 nm thick active layer of cells with 0.01% and 1 % doped p-type silicon oxide layer, (b) HIT cell with i-aSiO:H or 0.01% doped δp-aSiO:H as passivation layer to cSi(n)/p-aSi:H layer.

For a comparison the J-V characteristic curves of Cell-A and Cell-B are plotted together in Fig. 7(a) for 300 nm thick active layer. The comparison clearly shows that the device with 1% doped p-layer performs better than that of 0.01% doped p-layer. This is an expected outcome, and most of the reported solar cells use the 1% doped p-layer. However, it is to be noted that with 1% doping the optical transmission of the film decreases, leading to an expected reduction in current density of the device Although this result with 0.01% doped p-layer is not very good but it indicated that a certain high current density may also be obtained with a relatively low doped p-layer, indicating an ability of the 0.01% doped p-layer to enhance current density and voltage of a solar cell. We used this feature to see effect of the 0.01% doped p-layer on HIT-type solar cell.

### 3.6.    HIT-type solar cell

Using this characteristics we simulated HIT-type solar cell where the aSiO:H layer (at the cSi/p-aSi:H interface) and the 0.01% doped silicon oxide layers were separately used as passivation layers. The device structure used for this simulation is shown in Table II and the J-V characteristics of the device is shown in Fig. 7(b).

**Table III.** Solar cell device characteristics for simulated HIT-type solar cells with δp-aSiO:H (Cell-C) passivation layer or i-aSiO:H (Cell-D) passivation layer at the cSi/p-aSi:H interface.

| Cells | $V_{oc}$ (mV) | $J_{sc}$ (mA/cm$^2$) | FF (%) | PCE(%) | PmaxV (mV) | PmaxJ (mA/cm$^2$) |
|-------|-------|-------|-------|-------|-------|-------|
| Cell-C | 669.5 | 41.9 | 84.7 | 24.5 | 609.0 | 40.2 |
| Cell-D | 718.1 | 43.2 | 85.1 | 26.4 | 636.0 | 41.5 |

The device parameters for the Cell-C,D are shown in Table III. It shows that the device parameters improve with the use of the intrinsic silicon oxide layer in comparison to the 0.01% doped p-type oxide layer as the passivation layer. The primary  difference of the two passivation layers (intrinsic and delta-doped) are the difference in mid-gap and tail states. Passivation layer with a higher mid-gap state leads to poorer device performance. Therefore, it highlights the importance of using a less defective passivation layer to improve device performance.

## 4.    Conclusions

Evolution of band tail state and defect density was noticed on a lightly doped p-type silicon oxide material, estimated by constant photocurrent method. A sharp rise in dark conductivity and reduction in activation energy indicate efficient doping of the silicon oxide layer. However, rate of such a change reduces at a higher doping density. The CPM measurement indicates that a simultaneous rise in Urbach energy and midgap defect states in the lightly doped film. For higher level of doping these parameters are expected to increase further. It may be possible that the states detected by the CPM method is due to doping or desired active boron state, useful for device operation. Numerical simulation of device characteristics show an interesting device characteristics. In a HIT-type device it can provide additional built-in field to enhance device performance. In the HIT type device the improvement is thought to be due to the double roles played by the lightly doped p-aSiO:H film, that are the surface passivation as well as improvement in the valence and conduction band density of states.